\documentclass[1p,authoryear,12pt]{elsarticle}
\usepackage[utf8]{inputenc}
\usepackage[T1]{fontenc}
\usepackage{lmodern}
\usepackage[english]{babel}
\usepackage{amsmath}
\usepackage{amssymb}
\usepackage{graphicx}
\usepackage{wrapfig}
\usepackage{endnotes}

\let\footnote = \endnote

\begin{document}

\begin{frontmatter}
\title{Multifractal portrayal of the Swiss population}
\author{Carmen Delia VEGA OROZCO, Jean GOLAY, Mikhail KANEVSKI}
\address{Centre for Research on Terrestrial Environment, Faculty of Geosciences and Environment, University of Lausanne, 1015 Lausanne, Switzerland. Contact: CarmenDelia.VegaOrozco@unil.ch.}

\begin{abstract}
Fractal geometry is a fundamental approach for describing the complex irregularities of the spatial structure of point patterns. The present research characterizes the spatial structure of the Swiss population distribution in the three Swiss geographical regions (Alps, Plateau and Jura) and at the entire country level. These analyses were carried out using fractal and multifractal measures for point patterns, which enabled the estimation of the spatial degree of clustering of a distribution at different scales. The Swiss population dataset is presented on a grid of points and thus it can be modelled as a "point process" where each point is characterized by its spatial location (geometrical support) and a number of inhabitants (measured variable). The fractal characterization was performed by means of the box-counting dimension and the multifractal analysis was conducted through the Rényi's generalized dimensions and the multifractal spectrum. Results showed that the four population patterns are all multifractals and present different clustering behaviours. Applying multifractal and fractal methods at different geographical regions and at different scales allowed us to quantify and describe the dissimilarities between the four structures and their underlying processes. This paper is the first Swiss geodemographic study applying multifractal methods using high resolution data. 
\end{abstract}

\end{frontmatter}

\section{Introduction}
\label{Intro}
The spatial clustering of real patterns is an important subject in many fields and its characterization can be assessed by an ample number of indices \citep{Cressie93,Kanevski04,Illian08}. Here, clustering is defined as the spatial non-homogeneity of the way point patterns cover the geographical space in which they are embedded \citep{Tuia08}, and variability is related to the variation of the point density. Among these indices, fractal measures are widely developed. Their mathematical framework yields an useful tool for describing the irregularity or complexity of spatial phenomena and imparts great advantages over other traditional statistical methods.

Introduced by \citet{Mandelbrot67}, the word fractal was first coined to describe sets with abrupt and tortuous edges. A set of points whose any scale portion is statistically identical to the original object (statistical self-similarity) is fractal and it can be characterized by a fractal dimension which refers to the invariance of the probability distributions of the set under geometric changes of scale \citep{Rodriguez97}. In the case of multifractal point sets, all the moments of the probability distribution do not scale equivalently and an entire spectrum of generalized fractal dimensions is required \citep{Grassberger83b,Hentschel83,Paladin87,Tel89,Borgani93,Perfect06,Seuront10,Golay13}, i.e. the sparser and denser regions of a spatial distribution might have different scaling behaviours.

Many investigations have demonstrated that environmental, ecological and natural data are fractals. \citet{Bunde94} discussed in detail fractals in biology, chemistry and medicine. \citet{Burrough81} showed that many data of environmental variables and landscapes display a certain degree of statistical self-similarity over many spatial scales. \citet{Goodchild87} presented the relevance of fractals to geographic phenomena. \citet{Tucotte04} showed that landslides, forest fires and earthquakes presented a fractal distribution. \citet{Telesca01,Telesca04,Telesca06,Telesca07} characterized the spatial and temporal clustering behaviour of earthquake and forest fire sequences using fractal measures. \citet{Frontier87} and \citet{Seuront10} applied fractal theory to ecology and aquatic ecosystems.

In physical and human geography, fractal analyses have been carried out in many cases. A large literature on urban geography mentions the use of fractals to study the geometry and the creation of central places \citep{Arlinghaus85}, the town and city systems \citep{Francois95,Sambrook01}, the irregularities of city morphologies \citep{Batty94,Frankhauser94}, urban growth models \citep{Batty86,Batty89}, intra-urban built-up patterns \citep{Batty96,Frankhauser98,Keersmaecker03}, and the dynamics of population growth \citep{LeBras98,Ozik05}. \citet{Appleby96} applied multifractal methods to characterize the distribution pattern of the human population in the United States and Great Britain. \citet{Adjali01} analysed the multifractal behaviour of the distribution of human population in 10 countries around the world suggesting that the multifractal properties of their population distribution could be related to demographic and economic factors. These works have proved that the implementation of the fractal and multifractal formalism was relevant to urban studies.

In Switzerland, a fractal analysis has been carried out by (1) \citet{Frankhauser04} who compared the morphology of urban patterns in Europe; (2) \citet{Tannier13} who applied fractal measures to study the urban space structure and the delimitation of built-up areas in Basle; and (3) \citet{Kaiser09} who applied the lacunarity index for to study the clustering urban areas at different scales. Nonetheless, there are not known works concerning any structure analysis of the population distribution using multifractal measures at a local level.  

Thus, the present research aims at characterizing the spatial distribution pattern of the population in Switzerland at different scales through the fractal and multifractal formalism. The fractal dimension of the Swiss population distribution (SPD) was quantified using the box-counting method, while the multifractal dimensions were estimated using both Rényi's generalized dimensions and the multifractal spectrum. The population patterns in the three Swiss geographical regions (Alps, Plateau and Jura) were also studied separately and compared. These areas present different topographical features and different clustering behaviour. Therefore, by applying multifractal and fractal methods, we expected to quantify and depict their dissimilarities. Another innovation of this paper lies in the fact that the multifractal analysis of the population distribution was applied to high resolution data scaling from 250 m to 260 km, i.e. from an intra-city level to the country size with no aggregation of data at the city level as it has been done in the literature up to now.

Section 2 describes the dataset and the implemented methodology; section 3 provides the results and section 4 presents the conclusions of our findings.

\section{Theory/calculation}
\label{Theo}

\subsection{Data}
Switzerland is a landlocked country located in Western Europe. It borders France, Germany, Austria, Liechtenstein and Italy and it covers an area of 41,285 km$^{2}$. 

The census of the year 2000 counted 7,351,900 permanent residents and, since then, this amount has increased to 7,954,700 inhabitants \citep{FSO10}. The population in Switzerland has more than doubled since the beginning of the 20th century, starting from 3.3 million (1900) to 7.95 million (2011). In the period after World War II (1950-1970), the country underwent an important population growth with an annual average rate of about 1.4\%. It slowed down (0.6\%) from the 1970s to 1990s as a result of immigration restrictions and because of the economic recession. This growth was mostly concentrated in smaller centres and in agglomeration belts; while some larger urban centres experienced population decline. But, since then, the population growth rate has increased again to 0.8\% \citep{FSO10} while the population concentration has experienced a reversal trend. Nowadays, Switzerland can be considered as a densely populated country with an average population density of around 193 inhabitants per square kilometre.

\begin{figure}[t]
 \begin{center}
  \includegraphics[width=12cm]{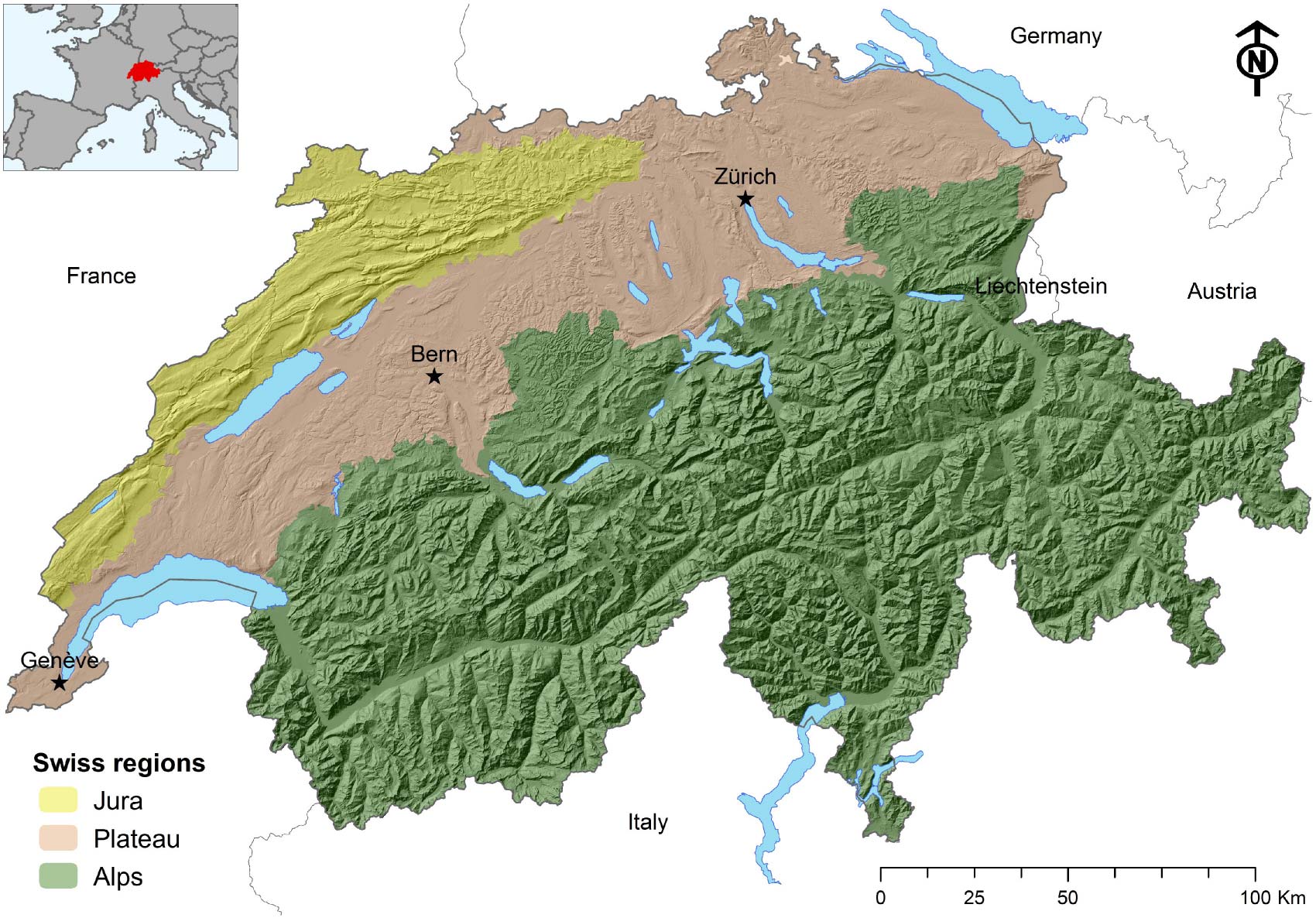}
 \end{center}
 \vspace{-20pt}
 \caption{Geographical regions in Switzerland}
 \label{Swiss}
\end{figure}

\begin{figure}[t]
 \begin{center}
  \includegraphics[width=13.5cm]{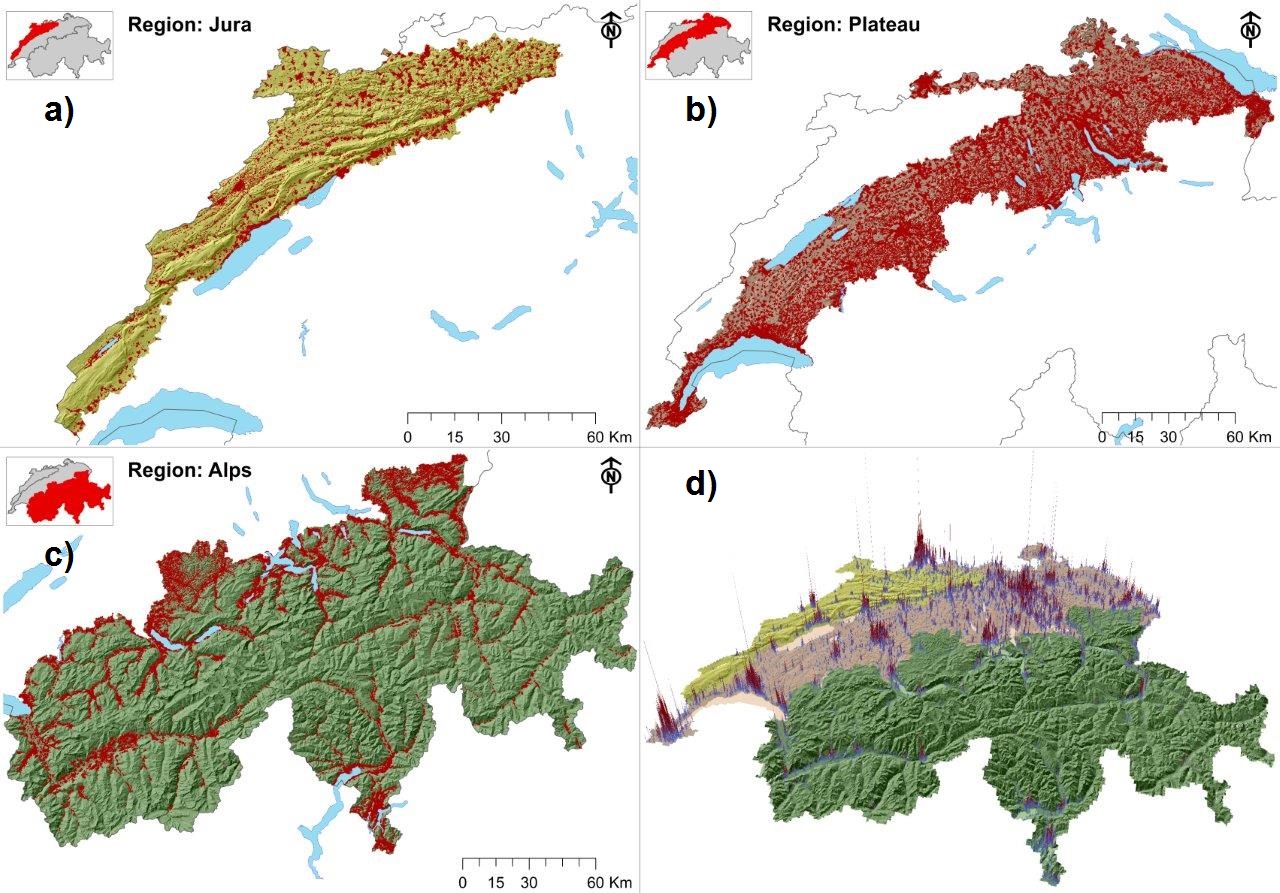}
 \end{center}
 \vspace{-20pt}
 \caption{Population distribution in Switzerland, year 2000. (a-c) Population distribution by geographic regions: Jura, Plateau and Alps, respectively. (d) 3D visualization of the $ln$-transformed data}
 \label{PoPRegion}
\end{figure}

Geographically, Switzerland is divided into three main regions: the Swiss Alps, the Plateau and the Jura, (see Figure \ref{Swiss}). Each region presents different geological and topographical features, and demographically, they clearly support dissimilar population distributions. Figure \ref{PoPRegion} displays the SPD of the year 2000 and a 3D visualization of the dataset where an inhomogeneous land-occupation structure is clearly detected with clusters of different sizes.

For instance, while the Alps occupy 60\% of the total country territory, only 23\% of the population lives in this highly mountainous region (average altitude of 1700 m). The Plateau, which is the economic epicentre, covers 30\% of the country's surface area and it concentrates $2/3$ of the total population, most of the Swiss industries and farmlands, as well as the major cities such as Geneva, Bern and Zurich. There are few regions in Europe that are more densely populated than the Plateau (450 people per km$^{2}$) and, in some areas such as the main cities, the population density can surpass 1000 people per km$^{2}$ \footnote{http://www.swissworld.org/en/geography/ consulted in 01/2013}. The Jura constitutes 10\% of the country and hosts 9\% of the population.

The population database used in the present study is the Swiss census of the year 2000. These high-resolution data are upheld and managed by the Swiss Federal Statistical Office \citep{FSO10} and they can be visualized through the 325,951 nodes (i.e. points) of a grid superimposed onto Switzerland, each of which is associated to the number of people living in an hectare (i.e.100 x 100 m).

\begin{figure}[t]
 \begin{center}
  \includegraphics[width=13.5cm]{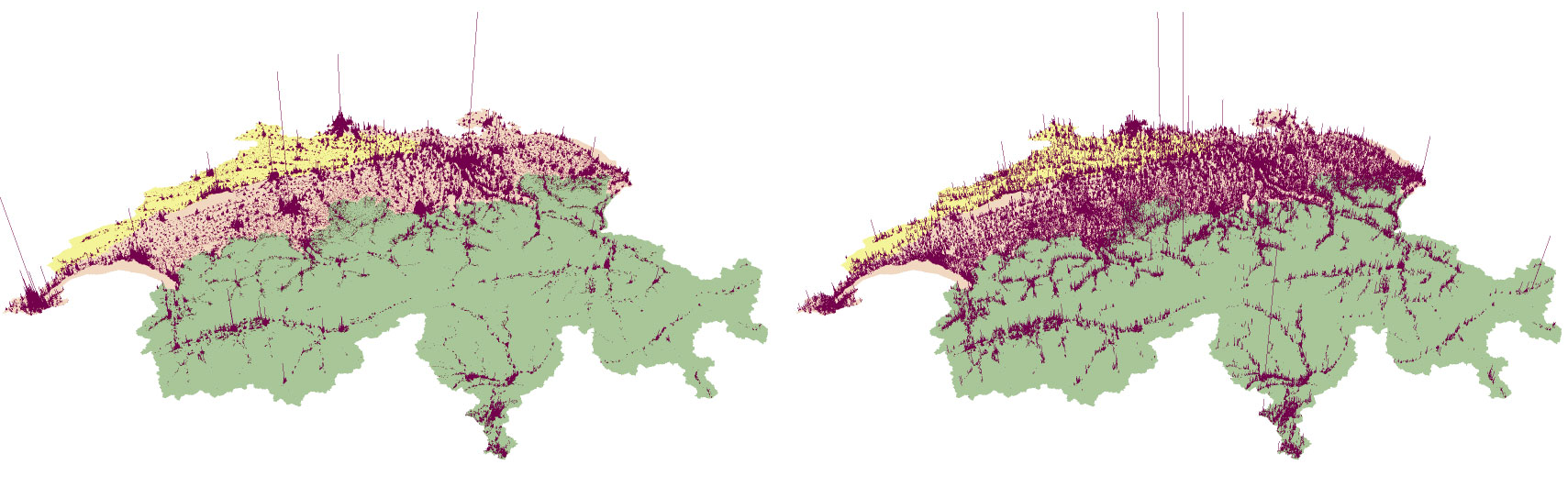}
 \end{center}
 \vspace{-20pt}
 \caption{The original population distribution in Switzerland (left) and one of the random permutation distribution simulated for comparisons (right)}
 \label{SimSwiss}
\end{figure} 

Reference patterns were generated to create confidence intervals of spatial randomness \citep{Illian08}. This procedure was done by creating a large number of random permutations ($N$=999) of the measured value (i.e. number of inhabitants) in order to destroy the dependence existing between the number of inhabitants and the spatial location of each grid node (or point). For this, the measured values were shuffled and then assigned randomly to the location points (or grid node) and the procedure was iterated $N$ times. See Figure \ref{SimSwiss}. The fractal and multifractal methods were also applied to these simulated samples and their results were compared with the original structure of the Swiss data to evaluate the departure of this raw pattern from a random structure.

\subsection{Methodology}
A point process is represented as sets of random points (events) generated within a space. Frequently, such processes exhibit a scaling behaviour indicating a high degree of point clustering over all scales \citep{Lowen95}. Fractal and multifractal tools can be used to characterize the intensity of spatial clustering at a wide range of scales \citep{Lowen95,Cheng95}.

\subsubsection{Fractal dimension}
According to \citet{Lovejoy86,Salvadori97,Tuia08}, a fractal dimension can be used to analyse the clustering properties (non-homogeneity) of point process realizations. If the studied distribution is embedded within a 2D space (e.g. geographical space), its fractal dimensions range from 0 (i.e. the topological dimension of a point) to 2 (i.e. the dimension of geographical space). If the points are dispersed or randomly distributed within the 2D study area, the corresponding fractal dimension is equal to 2; but this value decreases as the level of clustering increases and it can reach 0 if all the points are superimposed at one single location. Thus, fractal dimensions allow us to detect the appearance of clustering as a departure from a dispersed or random situation. 

A variety of fractal measures have been proposed such as the box-counting method \citep{Russell80,Lovejoy86,Tuia08}, the sandbox-counting method \citep{Grassberger83b,Daccord86,Feder88,Tel89,Vicsek90,Tuia08} and the information dimension \citep{Balatoni56,Hentschel83,Seuront10}. The fractal dimension of the SPD was estimated by applying the box-counting method.\\

\textit{The Box-Counting Method}

In the box-counting method, a regular grid of boxes of size $\delta$ is superimposed on the region under study and the number of boxes, $N(\delta)$, necessary to cover the point set is counted. Then, the linear size $\delta$ of the boxes is reduced and the number of boxes, $N$, is counted again. The algorithm goes on until a minimum size $\delta$ is reached. For a fractal pattern, the scales ($\delta$) and the number of boxes ($N(\delta)$) follow a power law:

\begin{equation}
 \label{dfbox} N(\delta) \propto \delta^{-df_{box}}
\end{equation}

where $df_{box}$ is the fractal dimension measured with the box–counting method \citep{Tuia08}. Theoretically, as introduced formerly, self-similarity is a property present over an infinite range of scales, however, in natural fractals, this property is only encountered statistically over a finite range of scales \citep{Abramenko2008} for which it is possible to consider -$df_{box}$ as the slope of the linear regression fitting the data of the plot which relates $log(N(\delta))$ to $log(\delta)$. 

\subsubsection{Multifractality}
As stated by \cite{Mandelbrot88}, the notion of self-similarity can be extended to measures (spreading mass or probability) distributed on an Euclidean support (e.g. a point set). In this context, fractal sets might be described by not just one fractal dimension, but rather by a function \citep{Stanley88} or a spectrum of interlinked fractal dimensions. Such fractal sets are said to be multifractal.

Two different approaches were used to conduct multifractal analysis: (1) Rényi's generalized dimensions \citep{Hentschel83,Grassberger83a,Paladin87,Tel89,Borgani93,Perfect06,Seuront10} and (2) the multifractal singularity spectrum \citep{Halsey86,Meakin86,Stanley88,Chhabra89}. These two methods are a generalization of the box-counting method. For both of them, a regular grid of boxes of size $\delta$ is superimposed on the point set and a normalized measure (probability distribution) is computed over all boxes \citep{Lopes09}.
 \\

\textit{Rényi's Generalized Dimensions}

The spectrum of generalized dimensions, $D_q$, is estimated by computing Rényi's information, $I_q(\delta)$, of $q_{th}$ order \citep{Renyi70}: 

\begin{equation}
 \label{RenyInfo} I_q(\delta) = \frac{1}{(1-q)} \ \log (\sum^{N(\delta)}_{i=1} p_i(\delta)^q)
\end{equation}

where $p_i=n_i/N$ is the probability mass function in the $i_{th}$ box of size $\delta$ and $q\in \mathbb{Z}$. Then, when applied to multifractal sets, $I_q(\delta)$ follows a power law:

\begin{equation}
 \label{RenyiPowerLaw} I_q(\delta) \propto \delta^{-D_q}
\end{equation}

Then, Rényi's generalized dimensions is defined as \citep{Hentschel83,Grassberger83a,Paladin87}:

\begin{equation}
 \label{GenDim} D_q = \lim_{\delta \to 0} \frac{I_q(\delta)}{\log(1/\delta)}
\end{equation}

The $D_q$ spectrum is obtained by the slope of the plot relating $log(I_q(\delta))$ to $log(1/\delta)$. For monofractal sets, $D_q$ is equal for all $q$ order moments, whereas, for multifractal sets, $D_q$ depends on $q$ and decreases as $q$ increases \citep{Hentschel83,Golay13} characterizing the variability of the measure ($p_{i}(\delta)$). For $q = 0$, all non-empty boxes are equally weighted and, consequently, $D_q$ is equivalent to the box–counting dimension, $df_{box}$, which corresponds to the dimension of the support. For $q > 0$, the mass within the boxes gradually gains more importance in the overall box contribution to Rényi's information. As a result, the larger the mass  within a box, the higher the weight of the box. Thus, higher $q$ order moments capture the scaling behaviour of regions where the mass is clumped. It is also interesting to notice that $D_1$ and $D_2$ correspond to the information dimension and the correlation dimension respectively \citep{Grassberger83b,Halsey86}.\\

\textit{Multifractal Singularity Spectrum}

The multifractal singularity spectrum describes the scaling behaviours of a measure through an interlinked set of Hausdorff dimensions, $f(\alpha)$, associated to a singularity strength $\alpha$ \citep{Halsey86,Meakin86,Chhabra89}. Let $p_i(\delta)$ be the probability in the $i_{th}$ box, then, the scales ($\delta)$ and $p_i(\delta)$ follow a power law:

\begin{equation}
 \label{alpha} p_i(\delta) \propto \delta^{\alpha_i}
\end{equation}  

where $\alpha_i$ is the singularity strength of the $i_{th}$ box of size $\delta$. Then, the number of boxes, $N_\alpha(\delta)$, having a singularity strength between $\alpha$ and $\alpha + d\alpha$, can be related to the box size, $\delta$, as:

\begin{equation}
 \label{N_Alpha} N_\alpha(\delta) \propto \delta^{-f(\alpha)}
\end{equation}

where $f(\alpha)$ is the Hausdorff dimension of the set of boxes with singularity strength $\alpha$ \citep{Halsey86,Meakin86,Chhabra89}. This singularity spectrum can be related to Rényi's generalized dimensions through a Legendre transform as:

\begin{equation}
 \label{Legendre} (q - 1) D_q = q\alpha(q) - f(\alpha(q))
\end{equation}

for more details see \citep{Halsey86,Meakin86,Mandelbrot88,Seuront10}. \citet{Chhabra89} proposed a method for determining the multifractal singularity spectrum as a function of the $q$ orders without the application of the Legendre transform. Let $\mu(q)$ be the normalized measure of the probabilities in the boxes of size $\delta$, such as:

\begin{equation}
 \label{Mu} \mu_i(q,\delta) = \frac{[p_i(\delta)]^q}{\sum_i [p_i(\delta)]^q}
\end{equation}

where, again, $q$ provides a tool for exploring denser and rarer regions of the singular measure \citep{Seuront10}. Then, $\alpha(q)$ and $f(\alpha(q))$ can be computed as:

\begin{equation}
 \label{Alpha_q} \alpha(q) = \lim_{\delta \to 0} \frac{\sum^{N(\delta)}_{i} \mu_i(q,\delta) \log p_i(\delta)}{\log \delta}
\end{equation}

and

\begin{equation}
 \label{F_Alpha_q} f(\alpha(q)) = \lim_{\delta \to 0} \frac{\sum^{N(\delta)}_{i} \mu_i(q,\delta) \log \mu_i (q,\delta)}{\log \delta}
\end{equation}

The multifractal spectrum is obtained by plotting the singularity spectrum $f(\alpha(q))$ vs. the singularity exponent $\alpha(q)$. For $q = 0$, $f(\alpha(0))$ takes its maximum value and is equal to $D_0$ and $df_{box}$. For $q = 1$, $f(\alpha(1)) = \alpha(1) = D_1$ is the information dimension.

\section{Results and discussion}
\label{ResDiss}

As it was exposed in section \ref{Theo}, the Swiss population can be divided into three subsets according to the main geographic regions: the Swiss Alps, the Plateau and the Jura. In this study, the analysis of the Swiss population was first conducted from a global perspective and was then narrowed down to each of the three subsets.  

\begin{figure}[t]
 \begin{center}
  \includegraphics[width=8cm]{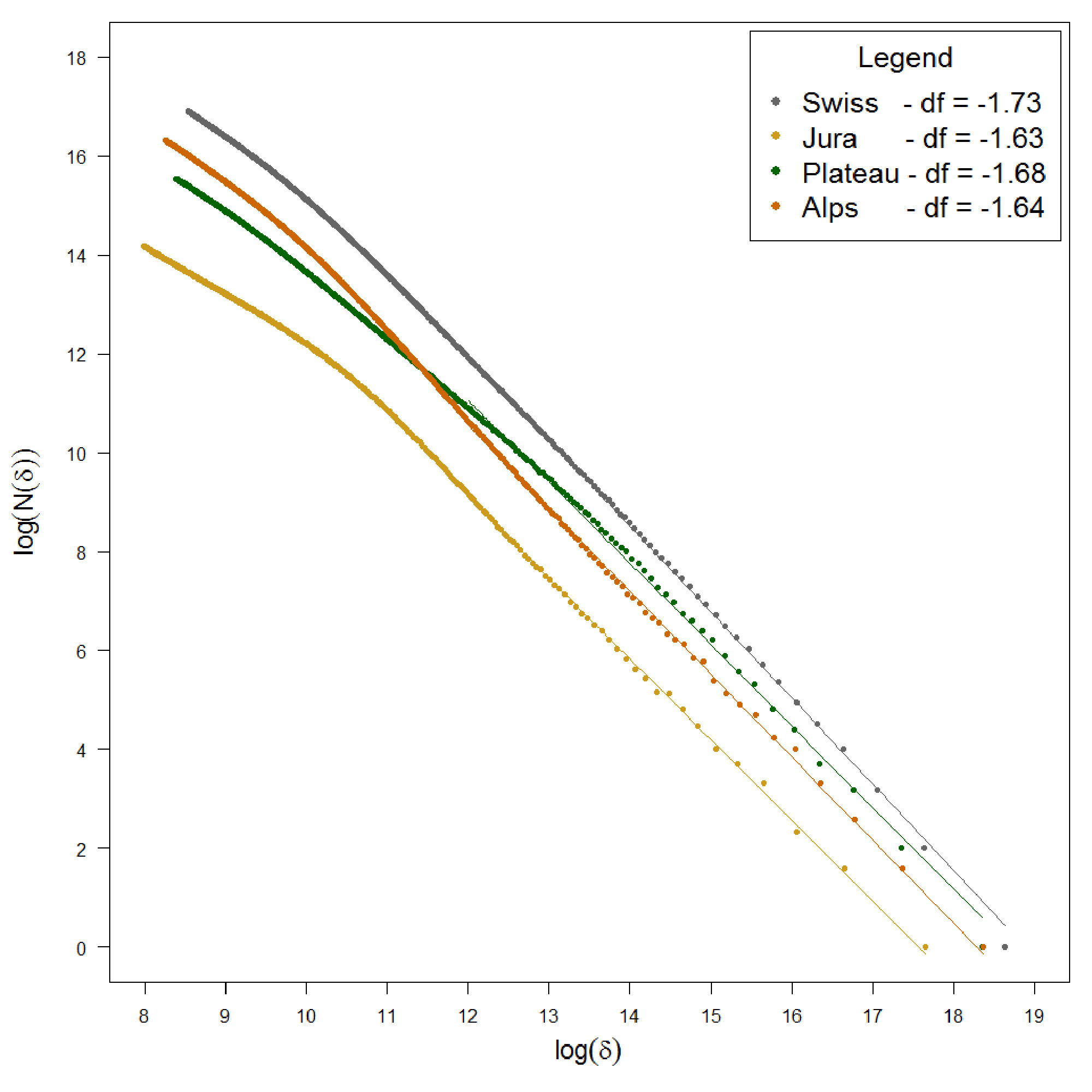}
 \end{center}
 \vspace{-20pt}
 \caption{Box-counting fractal dimension of the SPD in 2000. Scale range: 250 m $\leq \delta \leq$ 260 km}
 \label{BoxcountingPOP}
\end{figure}

Figure \ref{BoxcountingPOP} indicates the fractal dimension of the SPD in 2000 in both the entire country and the three geographic regions. It also exhibits the dependence between the logarithm of the Rényi information and the logarithm of the box size (250 m $\leq \delta \leq$ 260 km). These results bring to light three main behaviours characterizing different scale ranges. The first behaviour is detected at local scales $\leq$ 1 km ($log(\delta) =$ 10) and can be interpreted as a result of the influence of social-economic factors and/or territorial planning policies. These factors can be directly related to the way people concentrate in agglomeration areas. The second behaviour concerns scales ranging from 1 to $\sim$ 4 km ($log(\delta) =$ 12). It can be related to landuse practices. Finally, the third behaviour, which the rest of this paper will mainly focus on, is detected at scales greater than 4 km where topographic factors are likely to play a major role in constraining the way people appropriate their natural environment. 

The closer to 0 the values of $df_{box}$, the stronger the aggregation of the population, whereas, values of $df_{box}$ close to 2 indicate a homogeneous/random distribution. Comparing the $df_{box}$ values of the four patterns with the simulated samples reveals a significant deviation from an unstructured situation. Nevertheless, Figure \ref{BoxcountingPOP} does not reveal significant differences between the estimated $df_{box}$ values of the different regions, which leads to the conclusion that a comprehensive analysis of the SPD requires more than a single fractal dimension. 

To overcome the limitation of the fractal dimension in differentiating the four studied patterns, a multifractal analysis was carried out by means of Rényi's generalized dimensions, $D_q$, and the multifractal spectrum, $f(\alpha(q))$. 

\begin{figure}[t]
 \begin{center}
  \includegraphics[width=8cm]{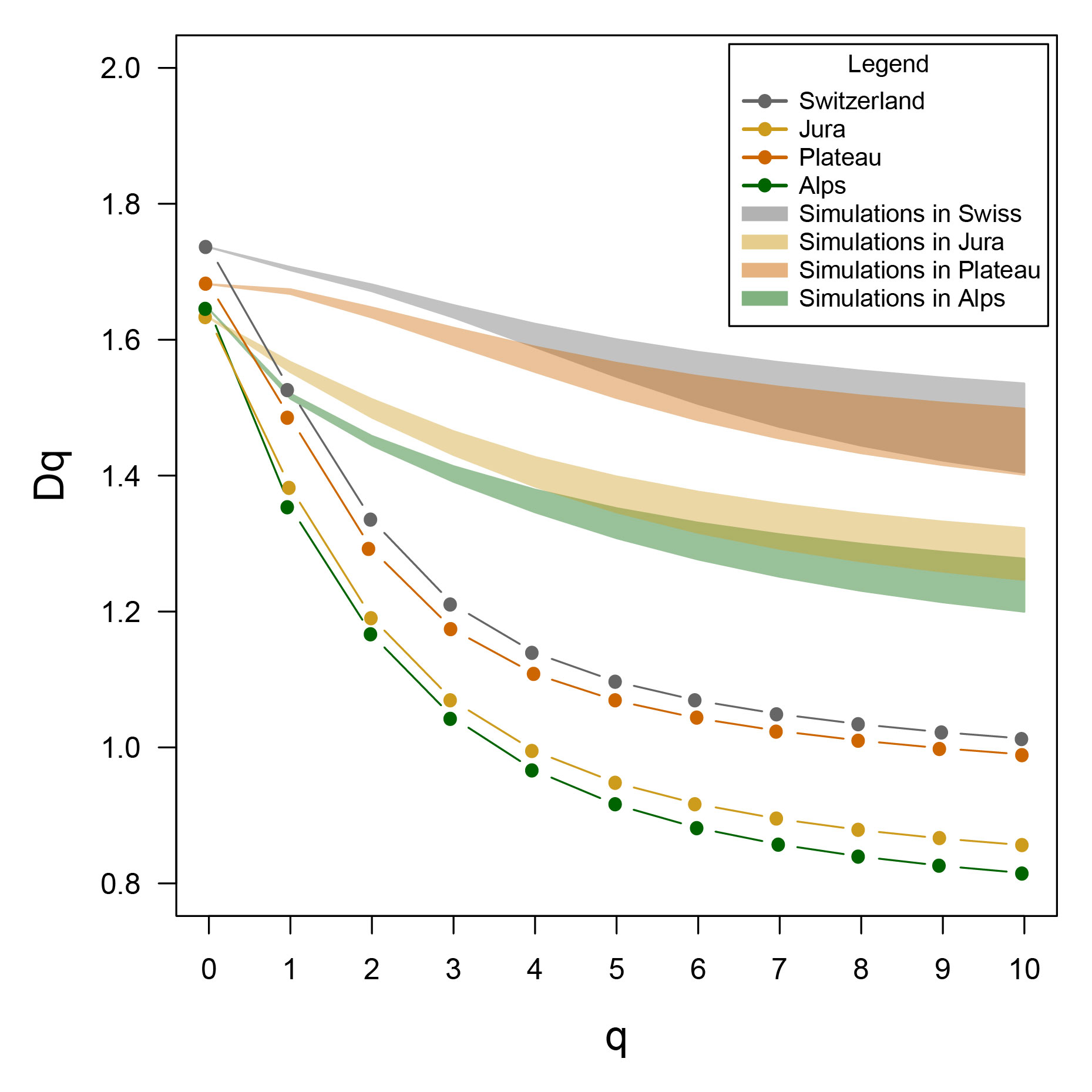}
 \end{center}
 \vspace{-20pt}
 \caption{Rényi's generalized dimensions of the SPD in 2000}
 \label{GenRenyDimPOP}
\end{figure}

Figure \ref{GenRenyDimPOP} shows Rényi's generalized dimensions for the four studied patterns and for $0\leq q \leq 10$. The dependence between $D_q$ and $q$ is non-linear for each considered SPD, which highlights their multifractal nature. $D_q$ values denote the degree of clustering of the distribution (non-homogeneity), whereas, the range width of the $D_q$ values ($D_{max}$ – $D_{min}$) is an indicator of the variability of the population density. 

The $D_q$ curves of the four SPD patterns decline faster than the simulated data, which indicates a substantial departure of the SPD patterns from shuffled/unstructured distributions. Furthermore, the $D_q$ of the SPD of the Alps and Jura regions decrease faster than that of both the Plateau and the entire country, meaning that the population in these regions (Alps and Jura) are more clustered. 

Regarding the range width of the $D_q$ values, the four curves revealed an extensive variability between the high and low densely populated areas for each of the considered regions, although it is more accentuated in the mountainous regions (Alps and Jura). This can also be depicted by comparing $D_0$, $D_1$ and $D_2$ values.

\begin{figure}[t]
 \begin{center}
  \includegraphics[width=8cm]{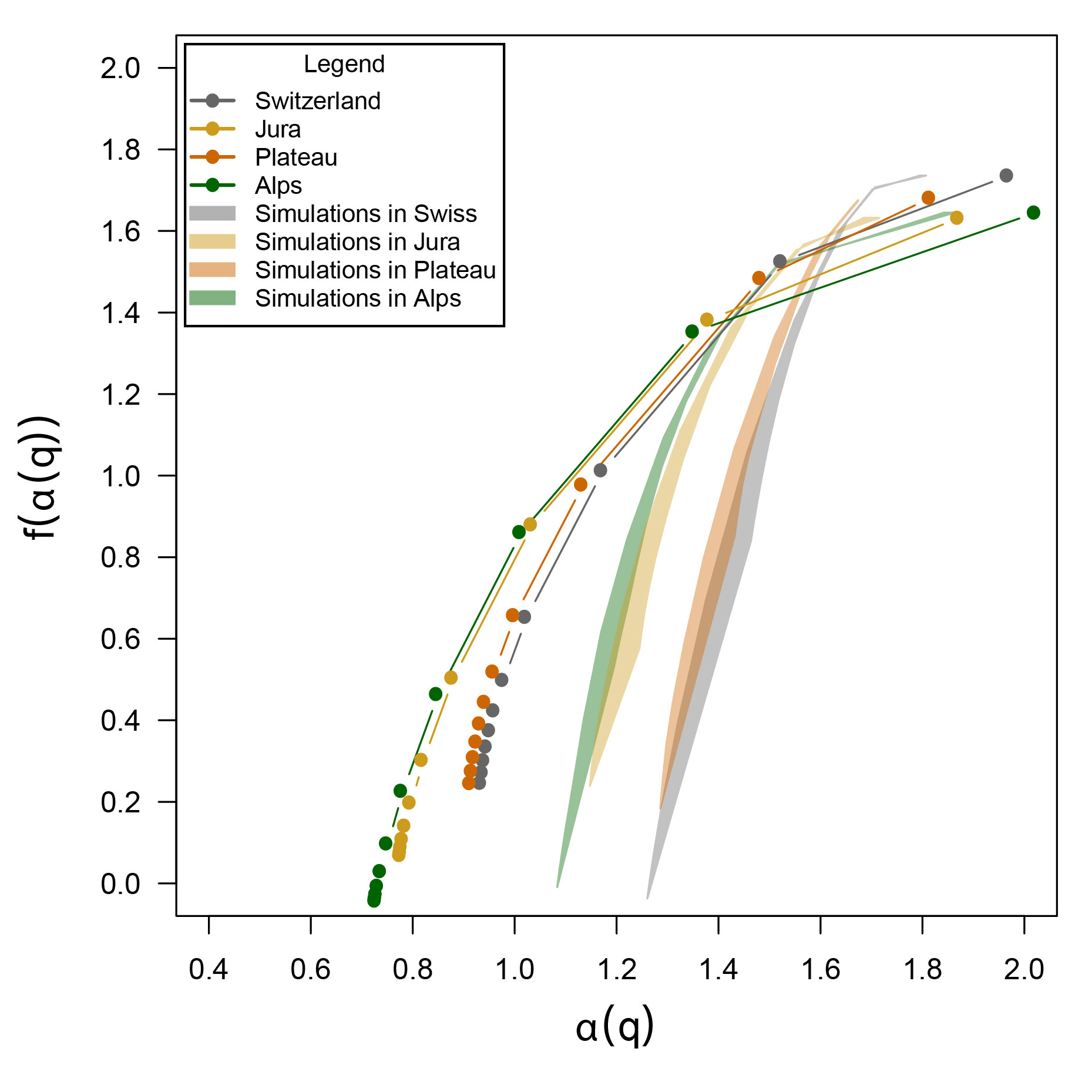}
 \end{center}
 \vspace{-20pt}
 \caption{The multifractal spectrum of the SPD in 2000}
 \label{MultiSpectPOP}
\end{figure}

The multifractal spectrum of the four SPD is illustrated in Figure \ref{MultiSpectPOP}. The multifractal spectrum of the four SPD is significantly different from that of the corresponding shuffled patterns. All the original distributions are characterized by an asymmetry skewed to the left with values lower than what is obtained for the shuffled data. This reflects the domination of large densely populated areas and highly clustered distributions.

The computed multifractal spectrum is in agreement with the estimations of Rényi's generalized dimensions. The SPD of the entire country has a multifractal behaviour similar to that of the population of the Plateau, but its heterogeneity is slightly higher as shown by the width of the $\alpha_{max}$ and $\alpha_{min}$ values. This similarity can be interpreted as a strong contribution of the the Plateau to the whole structure when analyses are conducted at entire country level. This follows from the fact that the Plateau concentrates two-thirds of the total population as well as the major cities of the country such as Zurich, Geneva, Basel, Bern and Lausanne.

The multifractal spectrum obtained for the Alps and the Jura are more skewed to the left than the previous ones, which is an indicator of highly clustered distributions. This is confirmed by the fact that their $f(\alpha(\delta))$ curves are lower and wider than that of both the Plateau and the entire country. Although, their multifractal natures are quite similar, the spatial structure of the population in the Alps is more heterogeneous than the Jura region.  

The significant clustering behaviour observed in the Alps and Jura regions can be mainly explained by the influence of the topography. Topographic features in these mountainous regions are the major factors which constrain land occupation in forcing populations to live in specific areas of the main valleys.           

\section{Conclusions}

A fractal and multifractal analysis of the population distribution in Switzerland was carried out. This investigation is the first Swiss geodemographic analysis which applies the multifractal formalism to high resolution data. This analysis comes out onto the characterization of the spatial structure of the population distribution for both the entire country and three geographical subregions (e.g. the Jura, the Plateau and the Alps). Rényi's generalized dimensions and the multifractal spectrum of the four spatial population distributions (SPD) showed different scaling behaviours between the high and low densely populated areas, which led to the differentiation of the four point patterns.

Analyses highlighted that the SPD of the Plateau region was more homogeneous than that of the Alps and Jura regions, especially in the case of densely populated areas. Additionally, the study pointed out the importance of multifractal analyses at different scale levels and in different geographic subregions, which helps discerning between distinct underlying processes. The distribution of the Swiss population has certainly been shaped by the socio-economic history and the complex geomorphology of the country. These factors have rendered a highly clustered, inhomogeneous and very variable population distribution at different spatial scales. Therefore, the analysis of the SPD is an interesting and challenging task and constitutes a fundamental approach for urban studies.

\section{Acknowledgments}
This work was partly supported by the Swiss National Science Foundation Project No. 200021-140658, "Analysis and Modelling of Space-Time Patterns in Complex Regions".

\theendnotes

\bibliographystyle{elsarticle-harv}
\bibliography{ReferencesPOP}

\begin{thebibliography}{59}
\expandafter\ifx\csname natexlab\endcsname\relax\def\natexlab#1{#1}\fi
\expandafter\ifx\csname url\endcsname\relax
  \def\url#1{\texttt{#1}}\fi
\expandafter\ifx\csname urlprefix\endcsname\relax\def\urlprefix{URL }\fi

\bibitem[{Abramenko(2008)}]{Abramenko2008}
Abramenko, V., 2008. Multifractal nature of solar phenomena. In: Wang, P.
  (Ed.), Solar physics research trends. Nova, New York, Ch.~2, pp. 95--136.

\bibitem[{Adjali and Appleby(2001)}]{Adjali01}
Adjali, I., Appleby, S., 2001. The multifractal structure of the human
  population distribution. In: Tate, N., Atkinson, P. (Eds.), Modelling scale
  in geographical information science. Wiley, Chichester, Ch.~4, pp. 69--85.

\bibitem[{Appleby(1996)}]{Appleby96}
Appleby, S., April 1996. Multifractal characterization of the distribution
  pattern of the human population. Geogr. Anal. 28~(2), 147--160.

\bibitem[{Arlinghaus(1985)}]{Arlinghaus85}
Arlinghaus, S., 1985. Fractals take a central place. Geograf. Annal. Series B,
  Hum. Geogr. 67~(2), 83--88.

\bibitem[{Balatoni and Rényi(1956)}]{Balatoni56}
Balatoni, J., Rényi, A., 1956. On the notion of entropy. Publ. Math. Inst.
  Hungarian Acad. Sci 1, 5--40, english translation in \textit{Selected Papers
  of Alfred Rényi}, vol. 1 (1976) 558-584, Akademiat Kiado, Budapest.

\bibitem[{Batty and Longley(1986)}]{Batty86}
Batty, M., Longley, P., 1986. The fractal simulation of urban structure.
  Environ. and Plan. A 18~(9), 1143--1179.

\bibitem[{Batty and Longley(1994)}]{Batty94}
Batty, M., Longley, P., 1994. Fractal cities. Academic Press, London.

\bibitem[{Batty et~al.(1989)Batty, Longley, and Fotheringham}]{Batty89}
Batty, M., Longley, P., Fotheringham, S., 1989. Urban growth and form: scaling,
  fractal geometry, and diffusion-limited aggregation. Environ. and Plan. A
  21~(11), 1447--1472.

\bibitem[{Batty and Xie(1996)}]{Batty96}
Batty, M., Xie, Y., 1996. Preliminary evidence for a theory of the fractal
  city. Environ. and Plan. A 28~(10), 1745--1762.

\bibitem[{Borgani et~al.(1993)Borgani, Plionis, and Valdarnini}]{Borgani93}
Borgani, S., Plionis, M., Valdarnini, R., February 1993. Multifractal analysis
  of cluster distribution in two dimensions. The Astrophys. J. 404~(1), 21--37.

\bibitem[{Bunde and Havlin(1994)}]{Bunde94}
Bunde, A., Havlin, S., 1994. Fractals in Science, 2nd Edition. Springer,
  Berlin.

\bibitem[{Burrough(1981)}]{Burrough81}
Burrough, P., November 1981. Fractal dimensions of landscapes and other
  environmental data. Nature 294~(5838), 240--242.

\bibitem[{Cheng and Agterberg(1995)}]{Cheng95}
Cheng, Q., Agterberg, F., 1995. Multifractal modeling and spatial point
  processes. Math. Geol. 27~(7), 831--845.

\bibitem[{Chhabra and Jensen(1989)}]{Chhabra89}
Chhabra, A., Jensen, R., March 1989. Direct determination of the f($\alpha$)
  singularity spectrum. Phys. Rev. Lett. 62~(12), 1327--1330.

\bibitem[{Cressie(1993)}]{Cressie93}
Cressie, N., 1993. Statistics for spatial data. Wiley, New York.

\bibitem[{Daccord et~al.(1986)Daccord, Nittman, and Stanley}]{Daccord86}
Daccord, G., Nittman, J., Stanley, H., January 1986. Radial viscous fingers and
  diffusion-limited aggregation: fractal dimension and growth sites. Phys. Rev.
  Lett. 56~(4), 336--339.

\bibitem[{De~Keersmaecker et~al.(2003)De~Keersmaecker, Frankhauser, and
  Thomas}]{Keersmaecker03}
De~Keersmaecker, M., Frankhauser, P., Thomas, I., October 2003. Using fractal
  dimensions for characterizing intra-urban diversity: the example of
  {B}russels. Geogr. Anal. 35~(4), 310--328.

\bibitem[{Feder(1988)}]{Feder88}
Feder, J., May 1988. Fractals ({P}hysics of solids and liquids). Plenum Press,
  New York.

\bibitem[{Frankhauser(1994)}]{Frankhauser94}
Frankhauser, P., 1994. La fractalité des structures urbaines. Anthropos,
  Paris.

\bibitem[{Frankhauser(1998)}]{Frankhauser98}
Frankhauser, P., 1998. The fractal approach: a new tool for the spatial
  analysis of urban agglomerations. Population 1, 205--240, an English
  Selection, Special issue \textit{New methodological Approaches in the Social
  Sciences}.

\bibitem[{Frankhauser(2004)}]{Frankhauser04}
Frankhauser, P., 2004. Comparing the morphology of urban patterns in europe: a
  fractal approach. In: Borsdorf, A., Zembri, P. (Eds.), European Cities
  Insights on Outskirts. Structures. Vol.~2. COST Action C10, Urban Civ. Eng.,
  Brussels, pp. 79--105.

\bibitem[{François et~al.(1995)François, Frankhauser, and
  Pumain}]{Francois95}
François, N., Frankhauser, P., Pumain, D., Juin 1995. Villes, densité et
  fractalité. nouvelles représentations de la répartition de la population.
  Annales de la recherche urbaine 67, 55--63.

\bibitem[{Frontier(1987)}]{Frontier87}
Frontier, S., 1987. Applications of fractal theory to ecology. In: Legendre,
  P., Legendre, L. (Eds.), Developments in Numerical Ecology. Springer Verlag,
  Berlin, pp. 335--378, nATO ASI Series G: Ecological Sciences Vol. 14.

\bibitem[{Golay et~al.(2013)Golay, Kanevski, Vega~Orozco, and
  Leuenberger}]{Golay13}
Golay, J., Kanevski, M., Vega~Orozco, C., Leuenberger, M., July 2013. The
  multipoint {M}orisita index for the analysis of spatial patterns. ArXiv
  preprint, 1--18.
\newline\urlprefix\url{http://arxiv.org/pdf/1307.3756.pdf}

\bibitem[{Goodchild and Mark(1987)}]{Goodchild87}
Goodchild, M., Mark, D., June 1987. The fractal nature of geographic phenomena.
  Annal. of the Assoc. of Am. Geogr. 77~(2), 265--278.

\bibitem[{Grassberger(1983)}]{Grassberger83a}
Grassberger, P., September 1983. Generalized dimensions of strange atractors.
  Phys. Lett. A 97~(6), 227--230.

\bibitem[{Grassberger and Procaccia(1983)}]{Grassberger83b}
Grassberger, P., Procaccia, I., October 1983. Measuring the strangeness of
  strange attractors. Phys. D 9~(1-2), 189--208.

\bibitem[{Halsey et~al.(1986)Halsey, Jensen, Kadanoff, Procaccia, and
  Shraiman}]{Halsey86}
Halsey, T., Jensen, M., Kadanoff, L., Procaccia, I., Shraiman, B., February
  1986. Fractal measures and their singularities: the characterization of
  strange sets. Phys. Rev. A 33~(2), 1141--1151.

\bibitem[{Hentschel and Procaccia(1983)}]{Hentschel83}
Hentschel, H., Procaccia, I., September 1983. The infinite number of
  generalized dimensions of fractals and strange attractors. Phys. D 8~(3),
  435--444.

\bibitem[{Illian et~al.(2008)Illian, Penttinen, Stoyan, and Stoyan}]{Illian08}
Illian, J., Penttinen, A., Stoyan, H., Stoyan, D., February 2008. Statistical
  analysis and modelling of spatial point patterns, 1st Edition. Wiley,
  Chichester, page 52.

\bibitem[{Kaiser et~al.(2009)Kaiser, Kanevski, Da~Cunha, and
  Timonin}]{Kaiser09}
Kaiser, C., Kanevski, M., Da~Cunha, A., Timonin, V., 2009. Emergence of swiss
  metropole and scaling properties of urban clusters. In: S4 International
  Conference on Emergence in Geographical Space, 23-25 November. S4, Paris, pp.
  23--25.

\bibitem[{Kanevski and Maignan(2004)}]{Kanevski04}
Kanevski, M., Maignan, M., 2004. Analysis and modelling of spatial
  environmental data. EPFL Press, Lausanne.

\bibitem[{Le~Bras(1998)}]{LeBras98}
Le~Bras, H., 1998. La planète au village: migrations et peuplement en France.
  Editions de l’Aube, France.

\bibitem[{Lopes and Betrouni(2009)}]{Lopes09}
Lopes, R., Betrouni, N., August 2009. Fractal and multifractal analysis: a
  review. Med. Image Anal. 13~(4), 643--649.

\bibitem[{Lovejoy et~al.(1986)Lovejoy, Schertzer, and Ladoy}]{Lovejoy86}
Lovejoy, S., Schertzer, D., Ladoy, P., January 1986. Fractal charaterization of
  inhomogeneous geophysical measuring networks. Nature 319~(6048), 43--44.

\bibitem[{Lowen and Teich(1995)}]{Lowen95}
Lowen, S., Teich, M., March 1995. Estimation and simulation of fractal
  stochastic point processes. Fractals 3, 183--210.

\bibitem[{Mandelbrot(1967)}]{Mandelbrot67}
Mandelbrot, B., May 1967. How long is the coast of {B}ritain? {S}tatistical
  self-similarity and fractional dimension. Science 156~(3775), 636--638.

\bibitem[{Mandelbrot(1988)}]{Mandelbrot88}
Mandelbrot, B., November 1988. An introduction to multifractal distribution
  functions. In: Stanley, H., Ostrowsky, N. (Eds.), Random fluctuations and
  pattern growth: experiments and models. Kluwer Academic, Dordecht, pp.
  279--291, nATO ASI Series E: Aplied Sciences Vol. 157.

\bibitem[{Meakin et~al.(1986)Meakin, Coniglio, Stanley, and Witten}]{Meakin86}
Meakin, P., Coniglio, A., Stanley, H., Witten, T., October 1986. Scaling
  properties for the surfaces of fractal and nonfractal objects: an infinite
  hierarchy of critical exponents. Phys. Rev. A 34~(4), 3325--3340.

\bibitem[{Office(2010)}]{FSO10}
Office, S. F.~S., 2010. Population, {P}anorama. Switzerland.

\bibitem[{Ozik et~al.(2005)Ozik, Hunt, and Ott}]{Ozik05}
Ozik, J., Hunt, B., Ott, E., October 2005. Formation of multifractal population
  patterns from reproductive growth and local resettlement. Phys. Rev. E
  72~(046213), 1--15.

\bibitem[{Paladin and Vulpiani(1987)}]{Paladin87}
Paladin, G., Vulpiani, A., 1987. Anomalous scaling laws in multifractal
  objects. Phys. Rep. 156~(4), 147--225.

\bibitem[{Perfect et~al.(2006)Perfect, Gentry, Sukop, and Lawson}]{Perfect06}
Perfect, E., Gentry, R., Sukop, M., Lawson, J., October 2006. Multifractal
  {S}ierpinski carpets: theory and application to upscaling effective saturated
  hydraulic conductivity. Geoderma 134~(3-4), 240--252.

\bibitem[{Rényi(1970)}]{Renyi70}
Rényi, A., 1970. Probability theory. Akadémiai Kiadò, Budapest.

\bibitem[{Rodríguez-Iturbe and Rinaldo(1997)}]{Rodriguez97}
Rodríguez-Iturbe, I., Rinaldo, A., 1997. Fractal river basins: chance and
  self-organization. Cambridge University Press, UK.

\bibitem[{Russell et~al.(1980)Russell, Hanson, and Ott}]{Russell80}
Russell, D., Hanson, J., Ott, E., October 1980. Dimension of strange
  attractors. Phys. Rev. Lett. 45~(14), 1175--1178.

\bibitem[{Salvadori et~al.(1997)Salvadori, Ratti, and Belli}]{Salvadori97}
Salvadori, G., Ratti, S.~P., Belli, G., 1997. Fractal and mutlifractal approach
  to environmental pollution. Environ. Sci. and Pollut. Res. 4~(2), 91--98.

\bibitem[{Sambrook and Voss(2001)}]{Sambrook01}
Sambrook, R., Voss, R., 2001. Fractal analysis of {US} settlement patterns.
  Fractals 9~(3), 241--250.

\bibitem[{Seuront(2010)}]{Seuront10}
Seuront, L., 2010. Fractals and multifractals in ecology and aquatic science.
  CRC Press, Boca Raton.

\bibitem[{Stanley and Meakin(1988)}]{Stanley88}
Stanley, H., Meakin, P., September 1988. Multifractal phenomena in physics and
  chemistry. Nature 335~(6189), 405--409.

\bibitem[{Tannier and Pumain(2013)}]{Tannier13}
Tannier, C., Pumain, D., 2013. Fractals in urban geography: a theoretical
  outline and an empirical example. Cybergeo: Eur. J. of Geogr. [on line],
  Syst., Model., GeostatDoc. 307. DOI: 10.4000/cybergeo.3275.

\bibitem[{Telesca et~al.(2007)Telesca, Amatucci, Lasaponara, Lovallo, and
  Rodrigues}]{Telesca07}
Telesca, L., Amatucci, G., Lasaponara, R., Lovallo, M., Rodrigues, M., October
  2007. Space-time fractal properties of the forest-fire series in central
  {I}taly. Commun. in Nonlinear Sci. and Numer. Simul. 12~(7), 1326--1333.

\bibitem[{Telesca et~al.(2001)Telesca, Cuomo, Lapenna, and
  Macchiato}]{Telesca01}
Telesca, L., Cuomo, V., Lapenna, V., Macchiato, M., January 2001. Identifying
  space-time clustering properties of the 1983-1997 {I}rpinia-{B}asilicata
  ({S}outhern {I}taly) seismicity. Tectonophys. 330~(1-2), 93--102.

\bibitem[{Telesca et~al.(2004)Telesca, Lapenna, and Macchiato}]{Telesca04}
Telesca, L., Lapenna, V., Macchiato, M., January 2004. Mono- and multi-fractal
  investigation of scaling properties in temporal patterns of seismic
  sequences. Chaos, Solitons and Fractals 19~(1), 1--15.

\bibitem[{Telesca and Lasaponara(2006)}]{Telesca06}
Telesca, L., Lasaponara, R., February 2006. Emergence of temporal regimes in
  fire sequences. Phys. A: Stat. Mech. and its Appl. 360~(2), 543--547.

\bibitem[{Tél et~al.(1989)Tél, F\"{u}l\"{o}p, and Vicsek}]{Tel89}
Tél, T., F\"{u}l\"{o}p, A., Vicsek, T., August 1989. Determination of fractal
  dimensions for geometrical multifractals. Phys. A: Stat. Mech. and its Appl.
  159~(2), 155--166.

\bibitem[{Tuia and Kanevski(2008)}]{Tuia08}
Tuia, D., Kanevski, M., May 2008. Envrionmental monitoring network
  charaterization and clustering. In: Kanevski, M. (Ed.), Advanced Mapping of
  Environmental Data: Geostatistics, Machine Learning and Bayesian Maximum
  Entropy. Iste/Wiley, London, Ch.~2, pp. 19--46.

\bibitem[{Turcotte and Malamud(2004)}]{Tucotte04}
Turcotte, D., Malamud, B., September 2004. Landslides, forest fires, and
  earthquakes: examples of self-organized critical behavior. Phys. A: Stat.
  Mech. and its Appl. 340~(4), 580--589.

\bibitem[{Vicsek(1990)}]{Vicsek90}
Vicsek, T., September 1990. Mass multifractals. Phys. A: Stat. Mech. and its
  Appl. 168~(1), 490--497.

\end{thebibliography}

\end{document}